\documentstyle[twocolumn,aps,epsfig, floats]{revtex}
\draft
\newcommand{\ket}[1]{\left | \, #1 \right \rangle}
\begin{document}

\twocolumn[\hsize\textwidth\columnwidth\hsize\csname
@twocolumnfalse\endcsname

\title{Experiments towards Falsification of Noncontextual Hidden Variable Theories }
\author{Markus Michler$^1$, Harald Weinfurter$^{2,3}$ and Marek \.Zukowski$^4$}
\address{$^1$Institut f\"ur Experimentalphysik,
 Universit\"at Innsbruck, A-6020 Innsbruck, Austria}
  \address{$^2$ Sektion Physik, Ludwig-Maximilians-Universit\"at, D-80799 M\"{u}nchen, Germany}
 \address{$^3$ Max-Planck-Insitut f\"{u}r Quantenoptik, D-85748 Garching, Germany}
 \address{$^4$Instytut Fizyki Teoretycznej i Astrofizyki Uniwersytet
  Gda\'nski, PL-80-952 Gda\'nsk, Poland} \date{Phys. Rev. Lett., {\bf 84}, 5457 (2000)}

\maketitle
\begin{abstract}
We present two experiments testing  the hypothesis of
noncontextual hidden variables (NCHV's). The first one  is based
on observation of two-photon pseudo-Greenberger-Horne-Zeilinger
correlations, with two of the originally three particles mimicked
by the polarization degree of freedom and the spatial degree of
freedom of a single photon. The second one, a single-photon
experiment, utilizes the same trick to emulate two particle
correlations, and is  an ``event ready" test of a Bell-like
inequality, derived from the noncontextuality assumption. Modulo
fair sampling, the data falsify NCHV's.
\end{abstract}


\vskip1pc

]

\narrowtext

The statistical nature of quantum predictions has frequently
initiated efforts to expand the quantum mechanical description.
The so-called hidden variable concepts try to cure the quantum
indeterminism by ascribing values to the properties of a system
which are defined already prior to the measurement. The question
arises whether such expansion of the theory is justified.

noncontextual hidden variable (NCHV) theories assume that the
predetermined result of a particular measurement does not depend
on what other observable is simultaneously measured (see e.g.
\cite{MERMIN90b}). Such cryptodeterminism was ruled out by the
Bell-Kochen-Specker (BKS) theorem \cite{BKS}, which, as a
mathematical theorem, does not need experimental confirmation --
and neither suggests one. Yet, quite recently, doubts arose about
the usefullness of the BKS theorem due to the impossibility of
experimentally testing the yes/no contradiction of the BKS theorem
in real world, where one is always confined to finite measurement
times and precision\cite{noBKS}. NCHV theories form a subset of
local realistic hidden variable (LHV) theories, which rely on the
more plausible assumption that the predetermined result of a
particular measurement does not depend on what other observable is
simultaneously measured in a spatially separated region. Bell's
theorem \cite{Bell} gives a clear prescription for a statistical
test of LHV theories. However, almost all experiments testing LHV
theories do not enforce spacelike separation of measurements, and
all are plagued with the low detection efficiency, thus falling
short of definitely invalidating LHV theories \cite{nospatialsep}.

We report two experiments testing the validity of NCHV theories.
The experiments are much simpler than equivalent Bell tests (with
no strict imposition of locality), and much less sensitive to
experimental imperfections. This includes lower threshold
interference visibility and consequently also less demanding
threshold for detection efficiencies. The possible adaptation to
other quantum systems paves the way to a loophole-free test of the
particular class of noncontextual hidden variable theories.
Formally, the experiments are employing the fact, that
measurements on distinct tensor product factors of Hilbert space
commute and can therefore form varying contexts for one another.
In the analysis of these experiments we obtain, for a specific
state \cite{PERES}, verifiable, statistical conditions for the
measurement results. In the first experiment the three particle
GHZ theorem \cite{GHZ,MerminGHZ}, by using its version for NCHV
theories \cite{ZUK}, is reduced to one with only two particles. In
the second one an ``event-ready" test of a Bell-like inequality
for only one particle allows to validate NCHV theories
\cite{HOME}.

If a NCHV theory attempts to reproduce quantum predictions it must
fulfill some basic prerequisites. First, the predetermined value
$a_i$, which is revealed when measuring the property
$\overline{A}$ for a given individual system $i$, i.e. a single
run of the experiment, must be equal to one of the eigenvalues of
the quantum mechanical observable $\hat{A}$ identified with this
property. Further, in quantum mechanics, any real function
$f(\hat{A},\hat{B})$ of commuting, and thus commeasurable
observables $\hat{A},$ $\hat{B}$ is also an observable, and the
eigenvalues of such a function observable are given by
$f({A},{B}), $ where ${A},$ ${B}$ are eigenvalues of the
respective operators. In a NCHV theory, this rule of functional
dependence must also hold for the preexisting values; i.e.,
$f(\overline{A},\overline{B})_i=f(a_i,b_i)$. We will see, that
there are input states to function observables for which NCHV
theories and quantum mechanics give conflicting predictions, and
which thus enable experimental tests.

The photon's momentum (and thus its propagation direction)
commutes, and thus is commeasurable with the photon's
polarization. We utilize this fact to define the observables in
our experimental test of NCHV theories. The first observable
$\hat{A}(\phi_A)$ is the direction of photon propagation behind
the (nonpolarizing) beam splitter BS (Fig. 1) and has eigenvalue
$A=+1$, if the photon is found  in the upper exit, and $A=-1$, if
in the lower one. The relative phase $\phi_A$ between the two
input paths is a free parameter of this observable and determines
the actual eigenstates.

The second observable, $\hat{B}(\phi_B)$, is the polarization of
the same photon. Its result $B=\pm1$ is determined operationally
by the exit port of a polarizing beam splitter (PBS), where the
photon is detected. The actual polarizations distinguished by the
PBS are set by a birefringent phase $\phi_B$. A third observable
$\hat{C}(\phi_C)$ acting on another photon is identical in its
nature to $\hat{B}(\phi_B)$.

\begin{figure}
\begin{center}
\epsfig{file=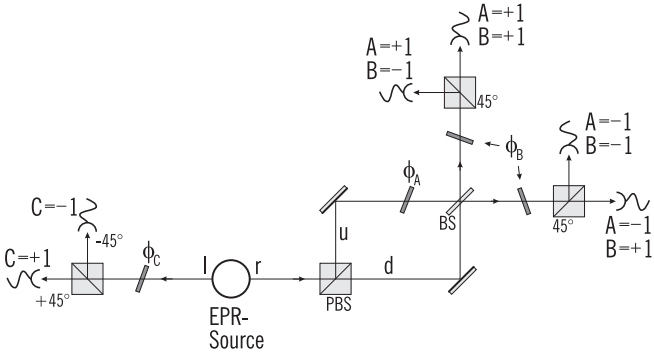,width=\linewidth}
\end{center}
\caption{Experimental set up to test noncontextual hidden variable
(NCHV) theories by determining correlations between the results of
commeasurable properties. These properties are the propagation
direction of a photon behind a beam splitter (BS), $\overline{A}$,
and the polarization of the photon, $\overline{B}$, as determined
by a polarizing beam splitter (PBS) oriented at 45$^\circ$ and the
birefringent phases $\phi_B$. The third observed property is the
polarization of a second photon, $\overline{C}$. Depending on the
relative phase $\phi_A$ between the inputs to the beam splitter
and on the birefringent phases $\phi_B$ and $\phi_C$, for certain
input states the results exhibit correlations which cannot be
described by NCHV theories. (For details of the actual layout see
text.)}
\end{figure}

Fig.\ 1 shows the experimental setup. Two polarization entangled
photons emerge from the source in the state
\begin{equation}
\ket{\psi_{initial}}=
\frac{1}{\sqrt{2}}\left(\ket{H}_2\ket{V}_1+\ket{V}_2\ket{H}_1\right)\ket{l}_2\ket
{r}_1, \label{1}
\end{equation}
 where $H$ and $V$ denote the
linear polarizations, $l$ and $r$ direction of propagation, and
the subscripts $1$ and $2$ enumerate the photons. After the
polarizing beamsplitter (PBS), which is oriented such that it
transmits H polarization and reflects V polarization, the state
changes into
 \begin{equation} \ket{\psi}=
\frac{1}{\sqrt{2}}\left(\ket{H}_2\ket{V}_1\ket{u}_1+\ket{V}_2\ket{H}_1\ket{d}_1
\right)\ket{l}_2, \label{2}
 \end{equation}
  where $u$ and $d$ denote
the two exit beams of the PBS. The factor in the bracket has the
formal structure of the GHZ state. The third particle is emulated
by an additional degree of freedom of particle $1$, which now is
also entangled with the polarization of particle $2$.

The beamsplitter (BS)  together with the phase shifter in front of
it  transforms the input modes $|u\rangle$ and $|d\rangle$ into
two output modes,
$\ket{\pm1,\phi_A}=\frac{1}{\sqrt{2}}(i\ket{d}_1\pm
e^{i\phi_A}\ket{u}_1)$. If detection of photon 1 happens behind
the upper exit of the BS we shall say that the eigenvalue $+1$ of
the dichotomic observable,  $\hat{A}(\phi_A)=
 \ket{+1,\phi_A}_{11}\langle +1,\phi_A|-\ket{-1,\phi_A}_{11}\langle
-1,\phi_A|$, was obtained, otherwise the  eigenvalue is $-1$.
 Detection behind one of the final PBS performs, together with the
 birefringent phase plate $\phi_B$ (with fast axis along V), a
 projection onto the states
 $\ket{\pm1,\phi_B}=\frac{1}{\sqrt{2}}(\ket{H}_1\pm
e^{i\phi_B}\ket{V}_1)$.
 Photon 2 is analyzed by the polarization observable
 $\hat{C}(\phi_C)$. For convenience of formal
 description, $\phi_C$ is realized by a birefringent plate with
 fast axis along H, i.e., $\ket{\pm1,\phi_C}=\frac{1}{\sqrt{2}}(\ket{V}_2\pm
e^{i\phi_C}\ket{H}_2)$. The formal definition of $\hat{B}(\phi_B)$
and $\hat{C}(\phi_C)$ follows the pattern of $\hat{A}(\phi_A)$.

The quantum  prediction for a photon prepared in state (2) to be
detected in one of the detectors which are assigned values
$A=\pm1$ and $B=\pm1$, and simultaneously the other photon to be
detected in one of the exits $C=\pm1$, is  given by
\begin{eqnarray} &P(A,B,C|\phi_A, \phi_B, \phi_C)& \nonumber \\
&= |(_1\langle A,\phi_A|_1\langle B, \phi_B|_2\langle C,
\phi_C|)|\psi\rangle|^2 & \nonumber \\&= \frac{1}{8}\left(1+ ABC
\sin{(\phi_A+\phi_B+\phi_C)}\right).&  \label{prob}
\end{eqnarray}
 The correlation function for the mean value of
the product of the three commeasurable observables monitored in
the first experiment is given by
\begin{eqnarray} &E(\phi_A,\phi_B,\phi_C)& \nonumber \\ &=\langle
\psi|\hat{A}(\phi_A)
\hat{B}(\phi_B)\hat{C}(\phi_C)\ket{\psi}=\sin{(\phi_A+\phi_B+\phi_C)}.&
\end{eqnarray}

In the second experiment, the registration of photon 2 serves as
trigger for the event-ready operation of the experiment on photon
1. After  the trigger event in the $45^\circ$ exit of the left
PBS, photon 1  propagates in the state
$\frac{1}{\sqrt{2}}(\ket{H}_1+\ket{V}_1)\ket{l}_1$, which changes
behind the right PBS into
 $ \ket{\psi'}_1=
\frac{1}{\sqrt{2}}(\ket{V}_1\ket{b}_1+\ket{H}_1\ket{a}_1 )$.
 The  correlation function for the
product of the observables $\hat{A}(\phi_A)$ and $\hat{B}(\phi_B)$
reads
\begin{eqnarray} &{\cal E}(\phi_A,\phi_B)=\langle \psi'|\hat{A}(\phi_A)
\hat{B}(\phi_B)\ket{\psi'}&\\ \nonumber&=\sin{(\phi_A+\phi_B)}.&
\nonumber \end{eqnarray}

Next, let us discuss the two experiments in terms of NCHV's. For
every emitted photon pair  the commuting observables now must have
preexisting values ${a}(\phi_A)_i$, ${b}(\phi_B)_i$, and
${c}(\phi_C)_i$, which are equal to $+1$ or $-1$. Context
independence here means that each of the values solely depends on
the associated phase, but not on the other two phase settings.
Therefore the correlation function must have the form:
\begin{equation} E_{NCHV}(\phi_A,\phi_B,\phi_C)=
\frac{1}{N}\sum_{i=1}^N{a}(\phi_A)_i{b}(\phi_B)_i{c}(\phi_C)_i ,
\end{equation}
 where $N$ is the (large) number of runs of the experiment.

Now, quantum mechanics predicts that the correlation functions
$E(0,0,\pi/2)$, $E(0,\pi/2,0)$, and $E(\pi/2,0,0)$ are all equal
to $1$. To reproduce such results the products of the preexisting
values $a(0)_ib(0)_ic(\pi/2)_i$, $a(0)_i b(\pi/2)_i c(0)_i$, and
$a(\pi/2)_i b(0)_i c(0)_i$ must therefore be all equal to $1$ in
every single run of the experiment.

From these three observations one obtains the NCHV prediction for
yet another set of phases. After multiplying the above three
products and using the fact that the square of any of the
preexisting values is one, it follows that the product $a(\pi/2)_i
b(\pi/2)_i c(\pi/2)_i$ is always equal to +1. However, the
resulting prediction, $E_{NCHV}(\pi/2,\pi/2,\pi/2)=1$, is in
absolute contradiction to the quantum mechanical prediction, for
which $E(\pi/2,\pi/2,\pi/2)=-1$.

In the second experiment, after the trigger detection of photon 2,
only observables $\hat{A}(\phi_A)$ and $\hat{B}(\phi_B)$ need to
be measured on the right side of the setup. For the preexisting
context independent values $a(\phi_A)_k$ and $b(\phi_B)_k$ the
Bell-type correlation function can be defined by
\begin{equation} {\cal E}_{NCHV}(\phi_A,\phi_B)=
\frac{1}{M}\sum_{k=1}^M{a}(\phi_A)_k{b}(\phi_B)_k, \end{equation}
where $k$ numbers the trigger events, and $M$ is their total
number. Such correlation functions must satisfy the well known
Bell-CHSH inequality
\begin{eqnarray} & -2\leq {\cal E}_{NCHV}(\phi_A,0) + {\cal
E}_{NCHV}(\phi_A,\pi/2)& \nonumber \\ & +{\cal
E}_{NCHV}(\phi_A',\pi/2)-{\cal E}_{NCHV}(\phi_A',0 )\leq 2.&
 \label{BELL}
  \end{eqnarray}
But quantum mechanic predicts values up to $2\sqrt{2}$.

For the experimental test of these contradicting predictions
polarization entangled photon pairs are produced by type-II
parametric down conversion and selected at a wavelength of 702nm.
In the right arm a polarizing beam splitter  was used to prepare
the state of Eq. (\ref{2}). Ideally, the polarizing beam splitter
transmits horizontal polarization and reflects vertical
polarization. In order to reduce residual H contributions in the
reflected beam $u$ (for our PBS about 4\%) and to obtain better
state preparation, we placed another PBS (not shown in Fig. 1),
rotated by 90$^\circ$ in this beam. The paths are combined at a
50:50 beam splitter which is insensitive to polarization.
 The phase
$\phi_A$ was locked to a dark fringe of a He:Ne interferometer
adjacent to the down conversion light.

The observables $\hat{B}(\phi_B)$ and $\hat{C}(\phi_C)$ are
measured behind Wollaston-type calcite prisms. The phases $\phi_B$
and $\phi_C$ are set with quartz plates, with the fast axis
oriented vertically for $\phi_B$ and horizontally for $\phi_C$.
The phases are determined by the birefringence of quartz and do
not need any stabilization on the time scale of the experiment
(about 50min).

Finally, after passing narrowband interference filters ($\Delta
\lambda = 5$~nm) the light was coupled to fiber pigtailed Silicon
single photon avalanche detectors. Only four detectors were
employed, i.e. the results $A=-1;B=-1$ and $A=-1;B=+1$ have not
been registered directly.

The symmetries of the experiment, particularly between the outputs
of the beam splitter BS, enable one to assume that the probability
to observe $(A=+1;B=+1;C=+1)$ is equal to the probability of the
result $(A=-1;B=-1 ;C=+1)$, and similarly for the other cases.
Thus the correlation function $E_{obs}(\phi_A, \phi_B, \phi_C)$
can be obtained from
\begin{displaymath}
E_{obs}(\phi_A, \phi_B, \phi_C)= 2 \sum_{B,C} (BC)P(1,B,C|\phi_A,
\phi_B, \phi_C).
\end{displaymath}

Fig.\ 2 shows the correlation functions for different settings of
the phases $\phi_B, \phi_C$, when varying the phase $\phi_A$
\cite{correct}. For demonstrating a noncontextual GHZ-like
contradiction one first has to establish the behaviour for the
phase settings $(\pi/2,0,0)$, $(0,\pi/2,0)$, and $(0,0,\pi/2)$.
From the raw data, i.e. not corrected for background, detector
efficiencies etc., we obtain the following values for the
correlation functions: $E_{obs}\left(0.46\pi, 0, 0 \right)=0.885$,
$E_{obs}\left(0.01\pi, \pi/2, 0 \right)=0.897$, and
$E_{obs}\left(0.01\pi, 0,  \pi/2 \right)=0.884$ \cite{phase}. With
coincidence rates of about 500~s$^{-1}$ and a sampling time of
10~s (see Fig. 3), the error for these correlation values amounts
to $\Delta E_{obs}=0.005$.

\begin{figure}
\begin{center}
\epsfig{file=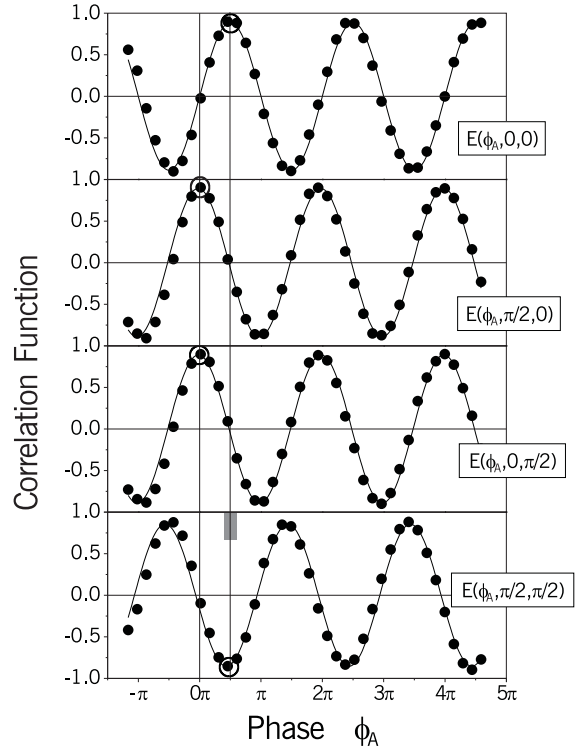,width=7.5cm}
\end{center}
\caption{GHZ-like correlation functions obtained in the first
experiment by varying phase $\phi_A$ for different settings of the
birefringent phases $\phi_B$ and $\phi_C$. From the encircled data
points of the first three correlation functions NCHV theories
deduce a prediction for the fourth correlation function which has
to lay within the shaded region. The contradiction with the data
is evident.}
\end{figure}

From these measurements contradicting predictions are deduced
within NCHV theories and quantum mechanics for the fourth
measurement with the setting $(\pi/2,\pi/2,\pi/2)$. The results of
Fig.\ 2 do not establish perfect correlations due to noise effects
like imperfect components and mode quality of the down conversion
fluorescence. Thus the contradiction between the two predictions
cannot be observed directly. The problem of less than perfect
correlations was already considered in the context of the GHZ
paradox. Similarily, within a NCHV description, the correlation
functions have  to fulfill the following inequality
\cite{MerminGHZ,GHZexp}:
\begin{eqnarray}&
2\ge E_{NCHV}(\phi_A,\pi/2,\pi/2)-E_{NCHV}(\phi_A,0,0)&\nonumber
\\ &
-E_{NCHV}(\phi_A',\pi/2,0)-E_{NCHV}(\phi_A',0,\pi/2)\ge  - 2  .&
\label{INEQ}
\end{eqnarray}

Thus, from the first three measurements NCHV theories predict for
the fourth measurement a lower bound of $
E_{NCHV}(0.46\pi,\pi/2,\pi/2) \geq 0.666 \pm 0.008$, which is in
strong disagreement with the observed value of $
E_{obs}(0.46\pi,\pi/2,\pi/2)=-0.885 \pm 0.005$.

\begin{figure}
\begin{center}
\epsfig{file=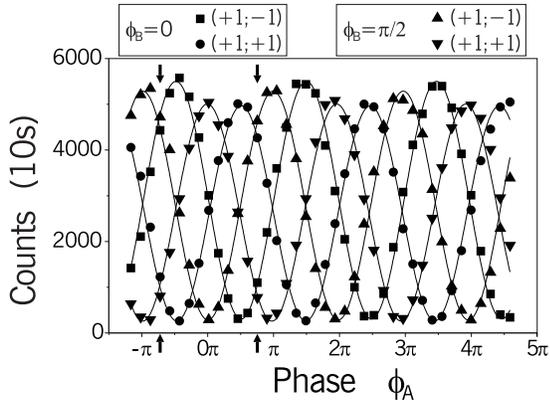,width=7.2cm}
\end{center}
\caption{Conditioned count rate at detectors on the right side
when varying the phase $\phi_A$ for two settings of $\phi_B$. The
data at the angles indicated by arrows lead to a violation of a
Bell-like inequality and thus again invalidate noncontextual
hidden variable theories.}
\end{figure}

In the second experiment detection of photon 2 at detector $C=+1$
with $\phi_C$ set to zero serves as a trigger to prepare photon 1
with a polarization of 45$^\circ$. One can pick up among the
conditioned counts (Fig.\ 3) observing $\hat{A}(\phi_A)$ and
$\hat{B}(\phi_B)$  the subset needed for calculating the
correlation functions and obtains, directly from the raw data,
$E_{obs}(-0.72 \pi,0)= 0.586 \pm 0.008$, ${E}_{obs}(-0.72
\pi,\pi/2)=0.705 \pm 0.008$, $E_{obs}(0.75\pi,0)=-0.590 \pm
0.008$, and $E_{obs}(0.75\pi,\pi/2)=0.714 \pm 0.008$ (the data
used are indicated by the arrows in Fig.\ 3). When evaluating the
Bell-like inequality (\ref{BELL}) with these values one obtains
$2.595 \pm 0.015 > 2$, i.e. we have again a gross violation of
noncontextuality.

The experiments had the feature that the effective collection
efficiency of photons was about 8\%. Therefore, like in all known
experiments testing the hidden variable theories, we have to
invoke the fair sampling assumption.

The present approach leads to radical simplification of the
experiments, which becomes possible when one concentrates on the
problem of noncontextuality.  Besides simpler setups and thus
reduced noise contributions, these schemes require less efficient
detection systems than standard Bell or GHZ experiments. For
perfect visibilities the required detection efficiency reduces to
$\sqrt{2}/2$. This value is still somewhat high for experiments
with entangled photon pairs, but surely can be reached by other
experimental techniques \cite{otherexp}. Not all hidden variable
theories can be ruled out with this approach, but the large class
of NCHV theories is now open to tests with single atoms or ions
bringing loophole-free tests into reach.

We acknowledge discussions with Abner Shimony and Anton Zeilinger.
This work was supported by the Austrian-Polish Program {\it
Quantum Information and Quantum Communication II} 11/98b, by the
UG Programs BW/5400-5-0264-9 and BW/5400-5-0032-0, by the Austrian
FWF (Proj. Y48-PHY), and the German DFG (Proj. WE2451/1), the last
stage by ESF Programme on {\it Quantum information theory and
quantum computation}.
 

\begin{references}
\bibitem{PERES} A. Peres  (Phys. Lett. A  {\bf 151},
107 (1990)), argues that the state independence  the the original
BKS paradox is only of an esthetic value. If there is no NCHV
description for a specific state, neither there is one for the
whole theory.
\bibitem{correct}In order to regain the sinusoidal shape of the
correlation function, for this figure only, the raw coincidence
rates are corrected for different overall detection efficiencies.
\bibitem{phase}Phases could be reproducibly set with an uncertainty of
$\Delta \phi_A = 0.01 \pi$ and $\Delta \phi_{B,C} = 0.005 \pi$.
\end{references}
\end{document}